\documentclass[prd,nofootinbib,showpacs,preprintnumbers,amsmath,amssymb]{revtex4}

\usepackage{graphicx}
\usepackage{dcolumn}
\usepackage{bm}

\def\bear{\begin{eqnarray}}
\def\ear{\end{eqnarray}}

\begin{document}

\title{Matrix Operator Approach to the Quantum Evolution Operator and the Geometric Phase}

\author{Sang Pyo Kim}\email{sangkim@kunsan.ac.kr}
\affiliation{Department of Physics, Kunsan National University, Kunsan 573-701, Korea}

\author{Jewan Kim}
\author{Kwang Sup Soh}
\affiliation{Department of Physics, Seoul National University, Seoul 151-742, Korea}

\medskip
\date{\today}

\begin{abstract}
The Moody-Shapere-Wilczek's adiabatic effective Hamiltonian and Lagrangian method is developed further into
the matrix effective Hamiltonian (MEH) and Lagrangian (MEL) approach to a parameter-dependent quantum system.
The matrix-operator approach formulated in the product integral (PI) provides not only a method to find
the wave function efficiently in the MEH approach but also higher order  corrections to the effective action systematically
in the MEL approach, a la the Magnus expansion and the Kubo cumulant expansion.
A coupled quantum system of a light particle of a harmonic oscillator is worked out, and
as a by-product, a new kind of gauge potential (Berry's connection) is found even for nondegenerate
cases (real eigenfunctions). Moreover, in the PI formulation the holonomy of the induced gauge
potential is related to Schlesinger's exact formula for the gauge field tensor. A superadiabatic
expansion is also constructed, and a generalized Dykhne formula, depending on the contour integrals
of the homotopy class of complex degenerate points, is rephrased in the PI formulation.

\noindent
Keywords: Matrix effective Hamiltonian, Matrix effective Lagrangian, Holonomy of nonabelian gauge potential, Superadiabatic expansion, Generalized Dykhne formula
\end{abstract}
\pacs{03.65.-w, 03.65.Vf, 03.65.Ca}

\maketitle

\section{Introduction}

There has been much interest in a quantum system depending on a set of parameters
(parameter-dependent system) since the geometric phase for cyclic evolution was discovered
by Berry \cite{berry84} and interpreted as a holonomy in a line bundle \cite{simon} or
a vector bundle \cite{aharonov-anandan,page} over the parameter space.
Since then, various approaches to the geometric phase have been introduced \cite{shapere-wilczek-book}.
There have appeared some methods to find the quantum evolution
operator (parameter-dependent wave function) in the superadiabatic basis \cite{berry87},
improving the accuracy of the adiabatic basis and taking into account the transitions
among the eigenstates of the Hamiltonian. An additional geometric factor responsible for
the transition amplitudes, known as the Dykhne formula \cite{dykhne}, was found for a
Hamiltonian system having complex degenerate points \cite{hwang-pechukas,MSW,berry90,JKP}.
Joye et al found a generalization of the Dykhne formula for a matrix Hamiltonian \cite{JKP}.

The matrix effective Hamiltonian (MEH) and the matrix effective Lagrangian (MEL) were initiated
by Moody et al \cite{MSW}. In the Hamiltonian approach, the quantum evolution
operator, expanded in the eigenstates of the Hamiltonian operator, becomes a matrix differential
equation \cite{MSW86,MSW}. In the Lagrangian approach, the full time-evolution kernel, expanded
in the basis of the Hamiltonian eigenstates, becomes an ordered matrix integration \cite{MSW}.
The MEL could be obtained by performing the ordered integration, but it has never been done due to
the noncommutativity of the matrix-valued operator on the parameter space; the formalism has not been
completed due to the lack of a method to handle the matrix-valued operators systematically.

On the other hand, independently of the MEH and the MEL approaches to a parameter-dependent
quantum system, a method formulated in the product integral (PI) \cite{dollard-friedman} was introduced
for the Wheeler-DeWitt (WDW) equation, a time-dependent Klein-Gordon like equation, frequently
encountered in quantum minisuperspace cosmological models, whose solutions were found
by explicitly using the properties of the PI \cite{kim91}. The positive and negative frequency decomposition
was done by using a method that can be regarded as a superadiabatic expansion method applied to relativistic quantum systems
\cite{kim92}.

In this paper, we shall develop further the Moody-Shapere-Wilczek's adiabatic effective method into
the MEH and the MEL approaches to a parameter-dependent quantum system.
The PI formulation provides us a powerful method to treat the matrix-valued operators and
a unified picture of the MEH, the MEL, geometric phases, the superadiabatic expansion and geometric
transition amplitudes. The matrix-operator approach formulated in the PI provides
not only a method to find wave functions efficiently in the MEH approach but also
higher-order corrections to the effective action systematically in the MEL approach, a la
the Magnus expansion \cite{magnus} and the Kubo cumulant expansion \cite{kubo}. The two-component wave function
and the effective action are worked out for a quantum system of a heavy particle
coupled to a light particle of a harmonic oscillator. In the MEL approach,
a new kind of gauge potential (Berry's connection)
is found even for nondegenerate eigenstates (real eigenfunctions). As a by-product of
the PI formulation, the holonomy of a nonabelian gauge potential
(Berry's connection) is observed to be related to Schleginger's formula for a gauge field tensor
\cite{schlesinger}, which is known as the nonabelian Stokes theorem \cite{bralic}.
Schleginger's formula is an exact formula for the geometric phase. A superadiabatic expansion
is also constructed, and a generalized Dykhyne formula, depending on the contour
integrals of the homotopy class of complex degenerate points, is rephrased in the PI formulation.

The organization of this paper is as follows: In Sec. II, the MEH for a parameter-dependent quantum
system will be found, and a gauge transformation for the wave function and a unitary
transformation for the Hamiltonian operator will be introduced that can separate
the matrix-valued vector gauge potential from the Laplacian operator into the diabatic term.
The MEH will be applied to a system of a heavy particle coupled to a light
particle of a harmonic oscillator, and the two-component wave function will be found in the
PI formulation. In Sec. III, the MEL will be found using the spectral resolution for a quantum system
consisting of subsystems interacting each other. The matrix effective action
will be constructed using the Magnus expansion and the Kubo cumulant expansion.
The full time-evolution operator kernel will be defined by using a path integral
with the matrix effective action. The MEL will be applied again to a heavy particle coupled to a light particle of the harmonic oscillator.
In Sec. IV, the geometric phase for a parameter-dependent Hamiltonian
will be reformulated as a holonomy of a matrix-valued gauge potential. The higher-order corrections to the Berry phase
will be considered a la the Magnus expansion. The holonomy of the
gauge potential will be related to Schlesinger's exact formula for the gauge field tensor.
In Sec. V, a method for the superadiabatic expansion will be introduced in the PI formulation.
In Sec. VI, the geometric transition amplitudes for a quantum system
having degenerate points coming from level crossing  of eigenvalues
will be shown to depend on the contour integrals of the homotopy class around the degenerate points.

\section{Matrix Effective Hamiltonian Approach}\label{sec-II}

\subsection{Formulation}

Let us consider a parameter-dependent quantum system whose time-independent Schr\"{o}dinger equation is given by
\begin{eqnarray}
\bigl[H_I (\vec{R}) + H_{II} (\vec{r}, \vec{R}) \bigr] \Psi (\vec{r}, \vec{R}) = E \Psi (\vec{r}, \vec{R}), \label{ham}
\end{eqnarray}
where the second Hamiltonian $H_{II}$ depends continuously on $\vec{R}$ as a parameter.
This kind of quantum system occurs frequently in many-body quantum systems consisting of
subsystems interacting each other. If $\Psi (\vec{r}, \vec{R})$ describes an atomic wave function
for which $ H_I$ is a sub-Hamiltonian for a nucleus (heavy particle) and
$H_{II}$ is another sub-Hamiltonian for an electron (light particle), then
the variable $\vec{r}$ is called the fast variable and the variable $\vec{R}$ the slow
variable. The method we shall use below depends on the assumption that there exists
a complete set of the eigenstates for $H_{II}$, allowing some
possible degeneracies corresponding to level crossings of $\vec{R}$-dependent eigenvalues.
This assumption is not too restrictive because the conditions for spectral resolution
can be generally satisfied in most quantum systems.

The Hamiltonian $H_{II}$ will be assumed to have a complete basis of orthonormal
eigenstates $\vert j \alpha (\vec{r}, \vec{R}) \rangle$,
\begin{eqnarray}
H_{II} (\vec{r}, \vec{R}) \vert j \alpha (\vec{r}, \vec{R}) \rangle = e_j (\vec{R}) \vert j \alpha (\vec{r}, \vec{R}) \rangle,
\label{eigen}
\end{eqnarray}
with the orthonormality
\begin{eqnarray}
\langle k \beta (\vec{r}, \vec{R}) \vert j \alpha (\vec{r}, \vec{R}) \rangle = \delta_{jk} \delta_{\alpha \beta},
\end{eqnarray}
where $j, k$ run for continuous as well as discrete indices and $\alpha, \beta$ account for possible degeneracies.
In this complete basis, the wave function $\Psi (\vec{r}, \vec{R})$ can be expanded as
\begin{eqnarray}
\Psi (\vec{r}, \vec{R}) = \sum_{j \alpha} \vert j \alpha (\vec{r}, \vec{R}) \rangle \psi_{j \alpha} (\vec{R})
= \Phi^T (\vec{r}, \vec{R}) \Psi(\vec{R}),
\end{eqnarray}
where we have introduced  a compact notation, arranging the basis $\vert j \alpha (\vec{r}, \vec{R}) \rangle$
as a column vector $\Phi (\vec{r}, \vec{R})$, and the amplitudes $\psi_{j \alpha} (\vec{R})$ as
another column vector $\Psi(\vec{R})$, have denoted the transpose of a matrix or a vector by $T$, and have assumed the
usual matrix operations.

Thus, the Schr\"{o}dinger equation may take the following matrix form:
\begin{eqnarray}
\bigl[\Phi^* (\vec{r}, \vec{R}) H_I (\vec{R}) \Phi^T (\vec{r}, \vec{R}) + \bar{H}_{II, D} (\vec{R}) \bigr] \Psi(\vec{R}) = E I \Psi(\vec{R}),
\label{MEH1}
\end{eqnarray}
where $*$ denotes the dual (bra) vectors, $I$ is the identity matrix, and $\bar{H}_{II, D} (\vec{R})$ is the diagonal matrix
\begin{eqnarray}
\bar{H}_{II, D} (\vec{R}) = \begin{pmatrix}
 e_1 (\vec{R}) & & & &   \\
& & \ddots & &\\
& & & e_j (\vec{R})&\\
& & & & \ddots
 \end{pmatrix}
\end{eqnarray}
arranged in the definite order used for the column vectors $\Phi (\vec{r}, \vec{R})$ and
$\Psi (\vec{R})$. The matrix operator on the left-hand side is a MEH. For a Hamiltonian $H_I$ of the form
\begin{eqnarray}
H_I (\vec{R}) = \frac{\vec{P}^2}{2M} + V_I (\vec{R}),
\end{eqnarray}
we introduce a matrix-valued gauge potential \cite{MSW,wilczek-zee} as
\begin{eqnarray}
\vec{A} (\vec{R}) = i \Phi^* (\vec{r}, \vec{R}) \nabla_{\vec{R}} \Phi^T (\vec{r}, \vec{R}), \label{gaug pot}
\end{eqnarray}
with components $\vec{A}_{j \alpha , k \beta} (\vec{R}) = i \langle j \alpha (\vec{r}, \vec{R}) \vert \nabla_{\vec{R}}
\vert k \beta (\vec{r}, \vec{R}) \rangle$. Then, the Schr\"{o}dinger equation (in the unit of $\hbar = 1$)
separates completely as
\begin{eqnarray}
\Bigl[- \frac{1}{2M} (I \nabla_{\vec{R}} - i \vec{A} (\vec{R}))^2 + I V_I (\vec{R}) + \bar{H}_{II, D} (\vec{R}) \Bigr] \Psi(\vec{R}) = E I \Psi(\vec{R}).
\label{MEH2}
\end{eqnarray}
The term $\bar{H}_{II, D} (\vec{R})$ represents a back-reaction of the system $H_{II}$ on the system $H_{I}$.
In the Born-Oppenheimer approximation in which one sets $\vec{A} (\vec{R}) = 0$, $\bar{H}_{II, D} (\vec{R})$ contributes to the effective potential
\begin{eqnarray}
V_{\rm eff} (\vec{R}) = I V_I (\vec{R}) + \bar{H}_{II, D} (\vec{R}).
\end{eqnarray}
However, this is not the only effect that the system $H_{II}$
has on the system $H_{I}$ because it also introduces the matrix-valued gauge potential to the kinetic energy term.
The induced gauge potential has an important quantum mechanical effect, that is, the Berry phase \cite{berry84,simon,aharonov-anandan,page,shapere-wilczek-book}.
Equation (\ref{MEH2}) has been rewritten in a compact form for the vector equation that was already derived in Refs. \cite{wilczek-zee,zygelman}.
What has been obtained hitherto is that the original parameter-dependent Schr\"{o}dinger equation
separates as a vector equation and that the Hamiltonian operator becomes a matrix operator called the MEH operator:
\begin{eqnarray}
\bar{H}_{\rm eff} (\vec{R}) = - \frac{1}{2M} (I \nabla_{\vec{R}} - i \vec{A} (\vec{R}))^2 + I V_I (\vec{R}) + \bar{H}_{II, D} (\vec{R}).
\label{eff ham}
\end{eqnarray}
More generally, the MEH operator takes the form
\begin{eqnarray}
\bar{H}_{\rm eff} (\vec{R}) =  \Phi^* (\vec{r}, \vec{R}) H_I (\vec{R}) \Phi^T (\vec{r}, \vec{R}) + \bar{H}_{II, D} (\vec{R}),
\end{eqnarray}
where the first term is a matrix of the expectation values with respect to the basis $\vert j \alpha (\vec{r}, \vec{R}) \rangle$.

We may separate the gauge potential from the Laplacian operator in the parameter $\vec{R}$-space by first
expanding the wave function in the complete basis and then inserting a gauge factor defined in terms of the PI:
\begin{eqnarray}
\Psi (\vec{r}, \vec{R}) = \Phi^T (\vec{r}, \vec{R}) \Bigl(\prod^{\vec{R}} \exp \bigl[i \vec{A}^* (\vec{R})
\cdot d\vec{R} \bigr] \Bigr) \tilde{\Psi}(\vec{R}). \label{PI}
\end{eqnarray}
Here, we have introduced a PI for the matrix-valued gauge potential in Eq. (\ref{gaug pot})
to replace the ordered integral that is a symbolic notation in quantum field theory (see Appendix \ref{app-a}); however, in our opinion the PI
should deserve much more attention than the ordered integrals in that the former has a more rigorous mathematical
foundation \cite{dollard-friedman} than the latter. Finally, the Schr\"{o}dinger equation becomes
\begin{eqnarray}
\Bigl[- \frac{1}{2M} I \nabla_{\vec{R}}^2 + I V_I (\vec{R}) + \Bigl(\prod^{\vec{R}} \exp \bigl[i \vec{A}^* (\vec{R}) \cdot d\vec{R} \bigr] \Bigr)^{-1}
\bar{H}_{II, D} (\vec{R}) \Bigl(\prod^{\vec{R}} \exp \bigl[i \vec{A}^* (\vec{R}) \cdot d\vec{R} \bigr] \Bigr) \Bigr] \tilde{\Psi}(\vec{R}) =
E I \tilde{\Psi} (\vec{R}).
\label{MEH3}
\end{eqnarray}
The gauge potential $\vec{A} (\vec{R})$ in Eq. (\ref{eff ham}) has been eliminated, and the matrix Laplacian operator
has become a diagonal-matrix Laplacian operator in Eq. (\ref{MEH3}), whereas the back-reaction of the system
$H_{II}$ on the system $H_{I}$ has gained an off-diagonal matrix (the last term), which is
known as a diabatic term in the literature \cite{smith}.
It should be noticed that Eq. (\ref{PI}) is a gauge transformation of the original wave function
and that Eq. (\ref{MEH3}) is a unitary transformation for the MEH. A similar
singular gauge transformation appears in the anyon system with the Chern-Simons action \cite{wilczek};
however, the similarity is expected naturally because the singular gauge transformation of the
anyon system is a kind of the gauge transformation, Eq. (\ref{PI}), with an abelian singular gauge potential $\vec{A} (\vec{R})$
and the PI becomes an ordinary integral:
\begin{eqnarray}
\prod^{\vec{R}} \exp \bigl[i \vec{A}^* (\vec{R}) \cdot d\vec{R} \bigr] = \exp \bigl[i \int^{\vec{R}} \vec{A}^* (\vec{R}) \cdot d\vec{R} \bigr].
\end{eqnarray}
In the anyon system, the boundary condition for the original single-valued wave function $\Psi (\vec{r}, \vec{R})$
is transferred into a different boundary condition for the gauge-transformed wave function
$\tilde{\Psi} (\vec{R})$, for example, such as the multi-valuedness.

The ultimate goal of the MEH approach to the quantum system, Eq. (\ref{ham}), is to find the energy
eigenvalues and the wave functions. To find the energy eigenvalues, we need to solve the MEH equation, Eq. (\ref{MEH3}),
which involves an unknown energy eigenvalues $E$. A typical method would be to expand the vector-valued
wave function $\tilde{\Psi} (\vec{R})$ in some basis of orthonormal functions of $\vec{R}$.
Then, the MEH equation becomes a blocked matrix eigenvalue problem. The wave function
can be obtained as an eigenvector; however, there is a more efficient method that rewrites the MEH equation, Eq.
(\ref{MEH3}), as a two-component equation of the first order. This method was successfully applied to find
the complete set of wave functions for the quantum Friedmann-Robertson-Walker (FRW) cosmology minimally coupled
to scalar fields in Refs. \cite{kim91,kim92}. The explicit exhibition of this method to determine
the wave functions will be done as an example.

\subsection{Application} \label{MEH ap}

As an application of the MEH approach to the quantum evolution operator, we shall consider the Hamiltonian
for a system of a heavy particle coupled to a light particle with a potential of the specific form
\begin{eqnarray}
H(\vec{r}, \vec{R}) = \frac{\vec{P}^2}{2M} + V_I (\vec{R}) + \frac{\vec{p}^2}{2m} + \frac{m}{2} \omega^2 (\vec{R})
\vec{r}^2. \label{mod ham}
\end{eqnarray}
Equation (\ref{mod ham}) is an effective Hamiltonian for two interacting particles because if $M$ is the mass of the
heavy particle and $m$ the mass of the light particle, then the potential terms describe a small perturbation
about the position $\vec{R}$. The super-Hamiltonian of the WDW equation for
the FRW cosmology minimally coupled to a free massive scalar field
has the form of Eq. (\ref{mod ham}), in which $R$ corresponds to the scale factor of the universe
while $r$ corresponds to the scalar field \cite{kim91,kim92}. The light particle of the Hamiltonian in Eq.
(\ref{mod ham}) describes a three-dimensional harmonic oscillator with a variable frequency
depending on the position $\vec{R}$ of the heavy particle.
The exact wave function and energy eigenvalue for the Hamiltonian  in Eq. (\ref{mod ham}) can be determined
by using a limiting process in which the light particle acts on the heavy particle through the back-reaction
and the gauge potential while the heavy particle acts on the light particle through the variable frequency and so on.

To simplify the model further without any loss of generality, we shall restrict ourselves to the two-dimensional problem
of a heavy particle coupled to the light particle of a harmonic oscillator:
\begin{eqnarray}
H(x, X) = \frac{P^2}{2M} + V_I (X) + \frac{p^2}{2m} + \frac{m}{2} \omega^2 (X)
x^2. \label{mod ham2}
\end{eqnarray}
The orthonormal eigenfunctions of Eq. (\ref{eigen}) are the harmonic wave functions
\begin{eqnarray}
\phi_j (x, X) = \Bigl(\frac{m \omega (X)}{\pi (2^j j!)^2} \Bigr)^{1/4} H_j (\sqrt{m \omega (X)} x)
\exp \Bigl(- \frac{1}{2} m \omega (X) x^2 \Bigr), \label{har wav}
\end{eqnarray}
where $H_j(x)$ are the Hermite polynomials. We further assume a positive frequency $\omega (X) > 0$, which guarantees the reality of the wave functions in Eq. (\ref{har wav}) and, thus, no level-crossings. Working with the number-state representation
and the creation and the annihilation operators acting on it is convenient. The creation and the annihilation operators
obviously depend on $X$ as a parameter:
\begin{eqnarray}
a^{\dagger} (X) &=& \sqrt{\frac{m \omega(X)}{2}} x - i \frac{1}{\sqrt{2 m \omega (X)}} p, \nonumber\\
a (X) &=& \sqrt{\frac{m \omega(X)}{2}} x +i \frac{1}{\sqrt{2 m \omega (X)}} p.
\end{eqnarray}
The number states $a^{\dagger} (X) a (X)$ also depend on $X$ as a parameter, and the harmonic wave functions in Eq. (\ref{har wav}) are
the coordinate representation of the number states. It is straightforward to see that the Hamiltonian of the
light particle in the number-state representation leads to the diagonal matrix
\begin{eqnarray}
\bar{H}_{II, D} (X) = \omega (X) \Bigl(a^{\dagger} (X) a (X) + \frac{I}{2} \Bigr), \label{har ener}
\end{eqnarray}
and that the matrix-valued gauge potential in Eq. (\ref{gaug pot}) becomes
\begin{eqnarray}
A(X) = i \frac{d \omega (X)/dX}{4 \omega(X)} \bigl[a^{2} (X) - a^{\dagger 2} (X) \bigr]. \label{har gaug}
\end{eqnarray}
We should remark that due to reality and no level-crossing of the eigenfunctions in Eq. (\ref{har wav}), the matrix-valued gauge potential
in Eq. (\ref{har gaug}) neither contains any diagonal part nor any singularity.
The operators occurring in Eqs. (\ref{har ener}) and (\ref{har gaug}),
\begin{eqnarray}
N_1 = a^{\dagger} (X) a (X) + \frac{I}{2}, \quad N_2 =  a^{\dagger 2} (X) + a^{2} (X), \quad N_3 = a^{\dagger 2} (X) - a^{2} (X),
\label{har al}
\end{eqnarray}
constitute a Lie algebra $so(2,1)$ with the basis $\{N_1, N_2 /2, N_3 /2 \}$ satisfying the commutators
\begin{eqnarray}
\Bigl[ N_1, \frac{N_2}{2} \Bigr] = 2 \frac{N_3}{2}, \quad \Bigl[ \frac{N_2}{2}, \frac{N_3}{2} \Bigr] = 2 N_1, \quad
\Bigl[ \frac{N_3}{2}, N_1 \Bigr] = - 2 \frac{N_2}{2}.
\label{har com}
\end{eqnarray}
In the number-state representation the operators in Eq. (\ref{har al}) represent constant matrices, and
the constancy of these operators will be understood throughout this paper.

With the aid of the commutators in Eq. (\ref{har com}), the diabatic (last) term in Eq. (\ref{MEH3}) turns out to be
\begin{eqnarray}
\tilde{H}_{II, D} (X) &=& \Bigl(\prod^{X} \exp \bigl[i A^* (X) \cdot dX \bigr] \Bigr)^{-1}
\bar{H}_{II, D} (X) \Bigl(\prod^{X} \exp \bigl[i A^* (X) \cdot dX \bigr] \Bigr) \nonumber\\
&=& \omega (X) \Bigl[\cosh\Bigl(\frac{d \omega (X)/dX}{\omega (X)} \Bigr) N_1 - \frac{1}{2} \sinh \Bigl(\frac{d \omega (X)/dX}{\omega (X)} \Bigr)N_2 \Bigr].
\label{diab}
\end{eqnarray}
Then, the MEH equation becomes
\begin{eqnarray}
\Bigl[- \frac{I}{2M} \frac{d^2}{dX^2} + I V_I (X) + \tilde{H}_{II, D} (X) \Bigr] \tilde{\Psi}(X) =
E I \tilde{\Psi} (X). \label{har MEH}
\end{eqnarray}
The diabatic term for the back-reaction of the light particle consists of both the diagonal entries (the first term) and
the off-diagonal entries (the second term) in Eq. (\ref{diab}).
Therefore, Eq. (\ref{har MEH}) is not the standard energy-eigenvalue problem. We may expand
the column vector wave function $\tilde{\Psi} (X)$ in some basis of orthonormal functions of $X$, for example, in the eigenfunctions for the
heavy particle, $H_{I} (X) = P^2/2M + V_I(X)$. Then, Eq. (\ref{har MEH}) turns out to be a blocked matrix energy-eigenvalue problem,
whose entries are again matrices of the same dimension as the eigenfunctions for $H_I (X)$,
which may be infinite dimensional. This causes a great difficulty in finding the energy eigenvalues for numerical purposes.
We may circumvent the infinite dimensionality problem by truncating the
blocked matrices to an appropriate size or by rearranging the double indices,
for instance, according to the sum of both indices.

Let us turn to the problem of finding the  wave function for the Hamiltonian in Eq. (\ref{mod ham2}).
We can make use of the eigenfunctions determined from the matrix equation, Eq. (\ref{har MEH}); in fact, this procedure involves
a double expansion by using the harmonic wave functions of $x$ and the eigenfunctions for Eq. (\ref{har MEH}).
Thus, the resultant wave functions may serve as a good approximation in some regions of the configuration space of $x$ and $X$.
These wave functions do not, however, show asymptotic forms.
Instead, we may employ the two-component wave functions \cite{kim91,kim92}. The final result is the two-component wave function
\begin{eqnarray}
\begin{pmatrix}
\Psi (x,X) \\
\frac{\partial \Psi (x, X)}{\partial X}  \end{pmatrix}
&=& \begin{pmatrix}
\Phi^T (x,X) & 0 \\
0 & \Phi^T (x, X) \end{pmatrix} \prod_{X_0}^{X} \exp \Bigl[ \begin{pmatrix}
i A^* (X) & I \\ 2M (I V_I (X) + \bar{H}_{II, D} (X) - EI) & i A^* (X)
\end{pmatrix} dX \Bigr] \nonumber\\ & &\times
\begin{pmatrix}
\int \Phi (x', X_0) \Psi (x',X_0) dx' \\
\int \Phi (x', X_0) \frac{\partial  \Psi (x', X_0)}{\partial X_0}dx' \end{pmatrix}. \label{har two-com}
\end{eqnarray}

What the two-component wave function implies is that once the energy eigenvalue is determined by some methods,
the wave function at $X$ is evolved from the wave function at $X_0$ according to Eq. (\ref{har two-com}).
Of course, the two-component wave function for the Hamiltonian in Eq. (\ref{mod ham2}) does not belong to a Cauchy initial-value
problem, but belongs to an elliptic-type eigenvalue problem. Nevertheless, there is still some similarity in finding the wave function between
the Hamiltonian in Eq. (\ref{mod ham2}) and the super-Hamiltonian for the FRW universe
minimally coupled to a free massive scalar field as mentioned earlier; the only difference is that
the latter has the opposite sign for the kinetic energy operator, indicating that the WDW
equation is of a hyperbolic type and that the two-component wave function is truly an evolution of the
Cauchy initial data.

To determine the wave function explicitly, one should be able to evaluate the PI in Eq. (\ref{har two-com})
in terms of the known algebraic operations. There is no general method to evaluate the PI in a closed form
except for a few cases. So we may use useful properties such as Eqs. (\ref{a-8}) and (\ref{a-10}) of the PI
to extract the dominant terms and to expand the remaining part in a perturbation series. Some methods to evaluate
the PI have been proposed in Refs. \cite{kim91,kim92} and used to find the wave function for the
FRW universe minimally coupled to scalar fields with various potentials.
According to the result of Ref. \cite{kim92}, the wave function in Eq. (\ref{har two-com}) should take the form
\begin{eqnarray}
\Psi (x,X) &=& \Phi^T (x,X) \bigl[P_1 (X) Q_{11} (X, X_0) + P_2 (X) Q_{21} (X, X_0)\bigr] {\rm Wr}^{-1} [P_1 (X_0), P_2 (X_0)]
\nonumber\\ && \times
\Bigl[ \frac{dP_2(X_0)}{dX_0} \int \Phi (x', X_0) \Psi (x',X_0) dx' - P_2 (X_0)\int \Phi (x', X_0) \frac{\partial  \Psi (x', X_0)}{\partial X_0}dx'   \Bigr]
\nonumber\\ && +
\Phi^T (x,X) \bigl[P_1 (X) Q_{12} (X, X_0) + P_2 (X) Q_{22} (X, X_0)\bigr] {\rm Wr}^{-1} [P_1 (X_0), P_2 (X_0)]
\nonumber\\ && \times
\Bigl[- \frac{dP_1(X_0)}{dX_0} \int \Phi (x', X_0) \Psi (x',X_0) dx' + P_1 (X_0)\int \Phi (x', X_0) \frac{\partial  \Psi (x', X_0)}{\partial X_0}dx'   \Bigr],
\label{har superad ex}
\end{eqnarray}
where $P_i (X)$'s are the fundamental solutions to the diagonal-matrix equation
\begin{eqnarray}
\Bigl[- \frac{I}{2M} \frac{d^2}{dX^2} + I V_I (X) + \bar{H}_{II, D} (X) - EI \Bigr] P_{1,2} (X) = 0,
\end{eqnarray}
${\rm Wr}^{-1} [P_1 (X_0), P_2 (X_0)]$ is the inverse of the Wronskian matrix, and
\begin{eqnarray}
 \begin{pmatrix}
Q_{11} (X, X_0) & Q_{12} (X, X_0) \\
Q_{21} (X, X_0) & Q_{22} (X, X_0) \end{pmatrix} = \prod_{X_0}^{X} \exp \Bigl[ \begin{pmatrix}
L_{11} (X) & L_{12} (X) \\ L_{21} (X) & L_{22} (X)
\end{pmatrix} dX \Bigr],
\end{eqnarray}
where
\begin{eqnarray}
L_{11} (X) &=&  {\rm Wr}^{-1} [P_1 (X), P_2 (X)] \Bigl[\frac{dP_2(X)}{dX} i A^* (X) P_1 (X) - P_2 (X) i A^* (X) \frac{dP_1(X)}{dX} \Bigr], \nonumber\\
L_{12} (X) &=&  {\rm Wr}^{-1} [P_1 (X), P_2 (X)] \Bigl[\frac{dP_2(X)}{dX} i A^* (X) P_2 (X) - P_2 (X) i A^* (X) \frac{dP_2(X)}{dX} \Bigr], \nonumber\\
L_{21} (X) &=&  {\rm Wr}^{-1} [P_1 (X), P_2 (X)] \Bigl[P_1 (X) i A^* (X)  \frac{dP_1(X)}{dX} - \frac{dP_1(X)}{dX}  i A^* (X) P_1 (X)\Bigr], \nonumber\\
L_{22} (X) &=&  {\rm Wr}^{-1} [P_1 (X), P_2 (X)] \Bigl[P_1 (X) i A^* (X)  \frac{dP_2(X)}{dX} - \frac{dP_1(X)}{dX}  i A^* (X) P_2 (X)  \Bigr].
\end{eqnarray}
The matrices $Q_{ij} (X, X_0)$ can be determined by using the Magnus expansion in Eq. (\ref{a-5}) or other methods, such as the $T$-ordered integral, Burum's method \cite{burum} and Salzman's method \cite{salzman}. This (superadiabatic)
expansion of the PI was used to find the asymptotic forms of the wave functions \cite{kim92}.

\section{Matrix Effective Lagrangian Approach}\label{sec-III}

\subsection{Formulation}

In the previous MEH approach to the quantum evolution operator, one should solve the vector equation, Eq. (\ref{MEH3}),
which is equivalent to the parameter-dependent quantum system in Eq. (\ref{ham}) through the spectral resolution
and the gauge transformation in Eq. (\ref{PI}). In order to use the MEH approach, however, one should surmount the difficult
task to find the exact solutions to Eq. (\ref{MEH3}). In this section, we shall develop the MEL approach,
the second half of the matrix-operator approach, to the quantum evolution operator.

Let us consider again the following Lagrangian for the quantum system in Eq. (\ref{ham}):
\begin{eqnarray}
L (\vec{r}, \vec{R}) = L_I (\vec{R}) + L_{II} (\vec{r}, \vec{R}).
\end{eqnarray}
Due to the parameter $\vec{R}$-dependence of the Lagrangian $L_{II}$, a common approach to the path
integral performs the $\vec{r}$-path integral first, whose result is an $\vec{R}$-dependent operator;
on the other hand, the $\vec{R}$-path integral might be very difficult to perform
in general without perturbation techniques applied to the $\vec{R}$-dependent operator. Contrary to
the path integral, in the MEL approach, we shall apply the spectral resolution of the system $II$
to the full time-evolution operator, the propagator in the configuration space
$(\vec{r}, \vec{R})$, and perform the remnant matrix-operator path integral for
the parameter $\vec{R}$-space. The MEL is a nonperturbative one associating the $\vec{R}$-dependent
operator with a matrix operator, which is not necessarily diagonal.

The full time-evolution operator kernel can be written as a Feynman sum over all the paths from
a configuration $(\vec{r}_i, \vec{R}_i)$ at time $t_i$ to another configuration $(\vec{r}_f, \vec{R}_f)$
at time $t_f$:
\begin{eqnarray}
K (\vec{r}_f, \vec{R}_f, t_f;  \vec{r}_i, \vec{R}_i, t_i)
= \int_{\vec{R}_i}^{\vec{R}_f} D[\vec{R}] \int_{\vec{r}_i}^{\vec{r}_f} D[\vec{r}]
\exp \Bigl[i \int_{t_i}^{t_f} ( L_I (\vec{R}) + L_{II} (\vec{r}, \vec{R}))dt \Bigr]. \label{ker}
\end{eqnarray}
Noticing that the total Lagrangian operator depends on $\vec{r}$ through $L_{II}$ only,
we may first perform the $\vec{r}$-path integral. The quantum state of the system $II$
in the basis of eigenstates of the Hamiltonian $H_{II}$ propagates as
\begin{eqnarray}
U_{II} (\vec{r}_f, \vec{R}_f, t_f;  \vec{r}_i, \vec{R}_i, t_i)
&=& \int_{\vec{r}_i}^{\vec{r}_f} D[\vec{r}]
\exp \Bigl[i \int_{t_i}^{t_f} L_{II} (\vec{r}, \vec{R})dt \Bigr] \nonumber\\
&=& \Phi^T (\vec{r}_f, \vec{R}_f) \prod_{t_i}^{t_f}
\exp \Bigl[- i \Bigl( \bar{H}_{II, D} (\vec{R}) + \vec{A}^T (\vec{R}) \cdot \dot{\vec{R}} \Bigr) dt \Bigr] \Phi^* (\vec{r}_i, \vec{R}_i).
\end{eqnarray}
The parameter $\vec{R}$ depends on $t$ implicitly so that $\dot{\vec{R}} = d \vec{R}/dt$. Then, the full time-evolution
operator kernel in the spectral resolution of  $H_{II} $ takes the form
\begin{eqnarray}
K (\vec{r}_f, \vec{R}_f, t_f;  \vec{r}_i, \vec{R}_i, t_i)
&=& \Phi^T (\vec{r}_f, \vec{R}_f) \int_{\vec{R}_i}^{\vec{R}_f} D[\vec{R}]
\exp \Bigl[i \int_{t_i}^{t_f} L_{I} (\vec{R}) dt \Bigr] \nonumber\\ && \times \prod_{t_i}^{t_f}
\exp \Bigl[- i \Bigl( \bar{H}_{II, D} (\vec{R}) + \vec{A}^T (\vec{R}) \cdot \dot{\vec{R}}
\Bigr) dt \Bigr] \Phi^* (\vec{r}_i, \vec{R}_i). \label{ker spec}
\end{eqnarray}
Next, we can factor out the diagonal matrix from the PI by using the general perturbation formula (sum rule) \cite{dollard-friedman}:
\begin{eqnarray}
\prod_{t_i}^{t_f}
\exp \Bigl[- i \Bigl( \bar{H}_{II, D} (\vec{R}) + \vec{A}^T (\vec{R}) \cdot \dot{\vec{R}}
\Bigr) dt \Bigr] = e^{- i \Omega (\vec{R}_f, \vec{R}_i)} \prod_{t_i}^{t_f}
\exp \Bigl[- i e^{i \Omega (\vec{R}, \vec{R}_i)} \vec{A}_{\rm O}^T (\vec{R}) \cdot \dot{\vec{R}}
e^{- i \Omega (\vec{R}, \vec{R}_i)} dt \Bigr],
\end{eqnarray}
where we have defined a frequency matrix as
\begin{eqnarray}
e^{- i \Omega (\vec{R}, \vec{R}_i)} = \exp \Bigl[- i \int_{t_i}^t \Bigl( \bar{H}_{II, D} (\vec{R}) + \vec{A}_{D}^T (\vec{R}) \cdot \dot{\vec{R}}
\Bigr) dt \Bigr],
\end{eqnarray}
with $\vec{A}_{D} (\vec{R})$ and $\vec{A}_{\rm O} (\vec{R})$ denoting the diagonal and the off-diagonal parts of
the matrix-valued gauge potential. This separation of the frequency matrix makes the Magnus expansion
of the ordered integrals converge in adiabatic regions, avoiding the possible divergence noticed in Ref. \cite{fernandez}.

In the MEL approach, we may write the full time-evolution operator kernel in Eq. (\ref{ker spec})
involving the PI by using an ordinary exponential operator a la the Magnus expansion (see Appendix \ref{app-a},
especially Eq. (\ref{a-5})):
\begin{eqnarray}
K (\vec{r}_f, \vec{R}_f, t_f;  \vec{r}_i, \vec{R}_i, t_i)
&=& \Phi^T (\vec{r}_f, \vec{R}_f) \int_{\vec{R}_i}^{\vec{R}_f} D[\vec{R}]
\exp \Bigl[i \int_{t_i}^{t_f} ( L_{I} (\vec{R}) - \bar{H}_{II, D} (\vec{R}) - \vec{A}_D^T (\vec{R}) \cdot
\dot{\vec{R}} )  dt \Bigr] \nonumber\\ && \times
\exp \Bigl[ \int_{t_i}^{t_f} M [- i e^{i \Omega (\vec{R}, \vec{R}_i)} \vec{A}_{\rm O}^T (\vec{R}) \cdot \dot{\vec{R}}
e^{- i \Omega (\vec{R}, \vec{R}_i)}] dt \Bigr] \Phi^* (\vec{r}_i, \vec{R}_i). \label{ker spec2}
\end{eqnarray}
The two matrix exponentials do not commute with each other; therefore, we need the Baker-Campbell-Hausdorff (BCH)
formula to get a single matrix exponential. However, the powerful cumulant expansion
method of Kubo \cite{kubo} extracts some pieces of information about the evolution operator without
the BCH formula. To obtain the diagonal-matrix element of the evolution operator, it suffices to find
the diagonal-matrix element of the second matrix exponential
\begin{eqnarray}
\Bigl< \exp \Bigl[ \int_{t_i}^{t_f} M [- i e^{i \Omega (\vec{R}, \vec{R}_i)} \vec{A}_{\rm O}^T (\vec{R}) \cdot \dot{\vec{R}}
e^{- i \Omega (\vec{R}, \vec{R}_i)}] dt \Bigr] \Bigr> =
\exp \Biggl[ \Bigl< \exp \Bigl[ \int_{t_i}^{t_f} M [- i e^{i \Omega (\vec{R}, \vec{R}_i)} \vec{A}_{\rm O}^T (\vec{R}) \cdot \dot{\vec{R}}
e^{- i \Omega (\vec{R}, \vec{R}_i)}] dt \Bigr] - 1 \Bigr>_c \Biggr], \label{cum}
\end{eqnarray}
where the average $ \langle \cdot \rangle$ is defined to take the diagonal element, and the subscript $c$
denotes the cumulant expansion of the matrix operator (see Appendix \ref{app-c}). Then, the matrix
effective action results from the diagonal part of the exponential operators in Eq. (\ref{ker spec2}):
\begin{eqnarray}
S_{\rm eff} (\vec{R}) = \int_{t_i}^{t_f} ( L_{I} (\vec{R}) - \bar{H}_{II, D} (\vec{R}) - \vec{A}_D^T (\vec{R}) \cdot
\dot{\vec{R}} )  dt  + \Bigl< \exp \Bigl[ \int_{t_i}^{t_f} M [- i e^{i \Omega (\vec{R}, \vec{R}_i)} \vec{A}_{\rm O}^T (\vec{R}) \cdot \dot{\vec{R}}
e^{- i \Omega (\vec{R}, \vec{R}_i)}] dt \Bigr] - 1 \Bigr>_c. \label{kubo eff act}
\end{eqnarray}
Finally, the full time-evolution operator kernel up to the diagonal-matrix effective action is approximated by
\begin{eqnarray}
K (\vec{r}_f, \vec{R}_f, t_f;  \vec{r}_i, \vec{R}_i, t_i)
= \Phi^T (\vec{r}_f, \vec{R}_f) \int_{\vec{R}_i}^{\vec{R}_f} D[\vec{R}]
\exp \Bigl[i S_{\rm eff} (\vec{R})  \Bigr] \Phi^* (\vec{r}_i, \vec{R}_i). \label{ker spec3}
\end{eqnarray}
We should point out that the time-evolution operator kernel using the matrix effective action is
not exact because the cumulant expansion of the matrix operator was undertaken with the diagonal-matrix
average in Eq. (\ref{cum}). Including the transition effects requires a transition matrix of the
form $I+T$, $T$ being an off-diagonal matrix that was not included in the diagonal-matrix average, to be inserted.

\subsection{Application} \label{MEL ap}

As an application of the MEL approach to the quantum evolution operator, we shall consider again
the system in Eq. (\ref{mod ham2}) whose Lagrangian is
\begin{eqnarray}
L(x, X) = \frac{M}{2} \dot{X}^2 - V_I (X) + \frac{m}{2} \dot{x}^2 - \frac{m}{2} \omega^2 (X) x^2. \label{har lag}
\end{eqnarray}
In order to find the full time-evolution operator kernel in Eq. (\ref{ker spec2}), we need to
perform the Magnus expansion
\begin{eqnarray}
\exp \Bigl[ \int_{t_i}^{t_f} M [- i e^{i \Omega (X, X_i)} A_{\rm O}^T (X) \dot{X}
e^{- i \Omega (X, X_i)}] dt \Bigr].
\end{eqnarray}
With the choice of the real eigenfunctions in Eq. (\ref{har wav}) for the light particle, the matrix-valued gauge
potential in Eq. (\ref{har gaug}) does not contain a diagonal part so that
\begin{eqnarray}
\Omega (X, X_i) = \int_{t_i}^{t} \omega(X) dt N_1, \quad A^T (X) = i \frac{d\omega(X)/dX}{4 \omega(X)} N_3.
\end{eqnarray}
Using the Lie algebra in Eq. (\ref{har com}) and the BCH formula, we can show that
\begin{eqnarray}
\exp \Bigl[ \int_{t_i}^{t_f} M [- i e^{i \Omega (X, X_i)} A_{\rm O}^T (X) \dot{X}
e^{- i \Omega (X, X_i)}] dt \Bigr] =
\exp \Bigl[ \int_{t_i}^{t_f} M [ \dot{X} (f_3 (X) N_3 + i f_2 (X) N_2)] dt \Bigr],
\end{eqnarray}
where
\begin{eqnarray}
f_3 (X) = \frac{d\omega(X)/dX}{4 \omega(X)} \cos \Bigl(2 \int_{t_i}^t \omega (X) dt \Bigr), \quad
f_2 (X) = \frac{d\omega(X)/dX}{4 \omega(X)} \sin \Bigl(2 \int_{t_i}^t \omega (X) dt \Bigr).
\end{eqnarray}

The first few terms of the Magnus expansion, Eq. (\ref{a-7}), are
\begin{eqnarray}
M_{(0)} &=& \dot{X} (f_3 (X) N_3 + i f_2 (X) N_2), \nonumber\\
M_{(1)} &=& -4 i \dot{X} \int_{X_i}^X dX_1 F_{[10]} (X_1, X) N_1, \nonumber\\
M_{(2)} &=& 4 i \dot{X} \int_{X_i}^X dX_1 \int_{X_i}^{X_1} dX_2  F_{[21]} (X_2, X_1)
(f_3 (X) N_2 + i f_2 (X) N_3) \nonumber\\ &&
- \frac{4}{3} i \dot{X} \int_{X_i}^X dX_1 \int_{X_i}^X dX_2  F_{[10]} (X_1, X)
(f_3 (X_2) N_2 + i f_2 (X_2) N_3), \nonumber\\
M_{(3)} &=& - 16 i \dot{X} \int_{X_i}^X dX_1 \int_{X_i}^{X_1} dX_2 \int_{X_i}^{X_2} dX_3  F_{[32]} (X_3, X_2)
(f_3 (X_1) f_3 (X)  + f_2 (X_1) f_2 (X)) N_1 \nonumber\\ &&
+ \frac{16}{3} i \dot{X} \int_{X_i}^X dX_1 \int_{X_i}^{X_1} dX_2 \int_{X_i}^{X_1} dX_3  F_{[21]} (X_2, X_1)
(f_3 (X_3) f_3(X) + f_2 (X_3) f_2(X)) N_1,
\end{eqnarray}
where
\begin{eqnarray}
F_{[ij]} (X_k, X_l) = f_2 (X_i) f_3 (X_j) - f_3 (X_i)  f_2 (X_j).
\end{eqnarray}
From the Lie structure in Eq. (\ref{har com}), the Magnus expansion must be closed in the form
\begin{eqnarray}
\exp \Bigl[ \int_{t_i}^{t_f} M [- i e^{i \Omega (X, X_i)} A_{\rm O}^T (X) \dot{X}
e^{- i \Omega (X, X_i)}] dt \Bigr] =
\exp \Bigl[ \int_{t_i}^{t_f} dt \dot{X} (G_1 (X) N_1 +  G_2 (X) N_2 + G_3 (X) N_3) \Bigr],
\end{eqnarray}
where $G_i$'s are determined by summing over $M_{(i)}$'s, that is,
\begin{eqnarray}
\dot{X} \sum_i G_i (X) N_i = \sum_{i} M_{(i)}.
\end{eqnarray}
Therefore, the full time-evolution operator kernel in Eq. (\ref{ker spec2}) becomes
\begin{eqnarray}
K (x_f, X_f, t_f;  x_i, X_i, t_i)
&=& \Phi^T (x_f, X_f) \int_{X_i}^{X_f} D[X]
\exp \Bigl[i \int_{t_i}^{t_f} ( L_{I} (X) - \omega (X) N_1 )  dt \Bigr] \nonumber\\ && \times
\exp \Bigl[ \int_{t_i}^{t_f} dt \dot{X} (G_1 (X) N_1 +  G_2 (X) N_2 + G_3 (X) N_3)
\Bigr] \Phi^* (x_i, X_i). \label{har ker spec2}
\end{eqnarray}
Here, $\Phi(x,X)$ represents the column vector of the harmonic wave functions in Eq. (\ref{har wav}).

The matrix effective action is obtained from Eq. (\ref{kubo eff act}) by evaluating the Kubo cumulant
expansion, Eq. (\ref{c-1}), in which the average $\langle \cdot \rangle$ is taken for the diagonal element.
After some algebra, we find the first few terms of the cumulant expansion to be
\begin{eqnarray}
&& \Bigl< \exp \Bigl[ \int_{t_i}^{t_f} dt \dot{X} (G_1 (X) N_1 +  G_2 (X) N_2 + G_3 (X) N_3)
\Bigr] -1 \Bigr>_c \nonumber\\ &&~~~~~~ = \int_{t_i}^{t_f} dt \dot{X} G_1 (X) N_1  + \Biggl[ \Bigl(
\int_{t_i}^{t_f} dt \dot{X} G_2 (X) \Bigr)^2 - \Bigl(
\int_{t_i}^{t_f} dt \dot{X} G_3 (X) \Bigr)^2 \Biggr] N_4 \nonumber\\ &&~~~~~~~~~
+ 2 \Bigl(\int_{t_i}^{t_f} dt \dot{X} G_1 (X) \Bigr) \Bigl(
\int_{t_i}^{t_f} dt \dot{X} G_2 (X) \Bigr)^2 N_4 + 2 \Bigl(
\int_{t_i}^{t_f} dt \dot{X} G_1 (X) \Bigr) \Bigl(
\int_{t_i}^{t_f} dt \dot{X} G_3 (X) \Bigr)^2 N_5 + \cdots,
\end{eqnarray}
where $N_i$'s are the constant matrices in the number-state representation for the
operators
\begin{eqnarray}
N_4 = a^{\dagger 2} (X) a^2 (X) + a^2 (X) a^{\dagger 2} (X), \quad
N_5 = a^{\dagger 2} (X) a^2 (X) - a^2 (X) a^{\dagger 2} (X).
\end{eqnarray}
Finally, up to the lowest Magnus expansion and the Kubo cumulant expansion, we obtain
the matrix effective action for the heavy particle as
\begin{eqnarray}
S_{\rm eff} = \int_{t_i}^{t_f} dt \Bigl[\frac{M}{2} \dot{X}^2 - g(X) N_1 \dot{X}
- V_I (X) - \omega (X) N_1  \Bigr], \label{har eff act}
\end{eqnarray}
where
\begin{eqnarray}
g(X) = \frac{1}{4} \int_{X_i}^{X} dX_1 \Bigl[ \frac{(d \omega(X_1)/dX_1) (d \omega(X)/dX)}{\omega(X_1) \omega(X)}
\sin \Bigl(2 \int^t \omega (X_1) dt - 2 \int^t \omega (X) dt \Bigr)\Bigr]. \label{har eff gaug}
\end{eqnarray}

If the canonical momentum $\pi = M \dot{X} - g(X)N_1$ is introduced for the heavy particle, the matrix effective Hamiltonian becomes
\begin{eqnarray}
H_{\rm eff} = \frac{1}{2M} \bigl(\pi + g(X) N_1 \bigr)^2
+ I V_I (X) + \omega (X) N_1. \label{har eff act2}
\end{eqnarray}
One can easily recognize that the effective action for the heavy particle involves a gauge potential
$-g(X) N_1$, as well as the back-reaction $\omega (X) N_1$ from the light particle.
The gauge potential results from a sequence of transitions accompanying the creation of
two quanta, followed by the annihilation of two quanta or vice versa, in the number-state
representation. Contrary to the common belief that the induced gauge potential does not exist
for the nondegenerate case (real eigenfunctions with no level-crossings), the MEL approach to the quantum evolution operator
shows that up to the lowest Magnus expansion and the Kubo cumulant expansion, the matrix
effective action has manifestly a new kind of mode-dependent gauge potential (Berry's connection).
The induced gauge potential $g(X) N_1$ cannot be removed and is thus physical because $N_1$ is invariant under a $U(1)$
gauge transformation of each eigenfunction via $e^{i \theta} a(X)$ and $e^{-i \theta} a^{\dagger}(X)$.

\section{Geometric Phase and Schlesinger's Formula}\label{sec-IV}

Let us consider a Hamiltonian $H(\vec{R})$ depending continuously on a parameter-vector
$\vec{R}$ in the parameter space which need not be of the specific form in Eq. (\ref{ham}),
but just depends on $\vec{R}$ as a parameter. In the spectral resolution, for each fixed $\vec{R}$,
the Hamiltonian satisfies the eigenvalue equation
\begin{eqnarray}
H(\vec{R}) \vert j \alpha (\vec{R}) \rangle = e_j (\vec{R})  \vert j \alpha (\vec{R}) \rangle, \label{eigen st2}
\end{eqnarray}
where the Latin letters $j, k, \cdots$ denote the energy levels and the Greek letters
$\alpha, \beta, \cdots$ denote the possible degeneracies of the Hamiltonian.
As the Hamiltonian $H(\vec{R})$ changes continuously along a curve $C(\vec{R})$,
the basis of the eigenstates $\vert j \alpha (\vec{R}) \rangle$ also changes continuously
along the curve, which has the analog of a moving coordinate system in classical mechanics.
We may arrange the eigenstates $\vert j \alpha (\vec{R}) \rangle$ as a column vector
$\Phi (\vec{R})$ as in Sec. \ref{sec-II} and introduce again the matrix-valued
gauge potential \cite{MSW,wilczek-zee} $\vec{A} (\vec{R})$ from Eq. (\ref{gaug pot}) in the form
\begin{eqnarray}
\nabla_{\vec{R}} \Phi (\vec{R}) = - i \vec{A} (\vec{R}) \Phi (\vec{R}). \label{gaug pot2}
\end{eqnarray}
Equation (\ref{gaug pot2}) is one of the Cartan's structure equations for a vector
bundle over the $\vec{R}$-space, which is the torsionless condition relating the change
of the orthonormal basis $\Phi (\vec{R})$ to the matrix-valued connection
$\vec{A} (\vec{R})$ \cite{EGH}. Under a unitary transformation, $\tilde{\Phi} (\vec{R}) = U(\vec{R})
\Phi (\vec{R})$,  corresponding to another choice of basis, the gauge potential can be shown to transform truly as a gauge field,
\begin{eqnarray}
\vec{\tilde{A}} (\vec{R}) = U(\vec{R}) \vec{A} (\vec{R}) U^{\dagger} (\vec{R})
+ i \bigl(\nabla_{\vec{R}}U(\vec{R}) \bigr) U^{\dagger} (\vec{R}),
\end{eqnarray}
rather than as a tensor. It is straightforward to see from the orthonormality of the basis
that the gauge potential is Hermitian, $\vec{A}^{\dagger} (\vec{R}) = \vec{A} (\vec{R})$.
The gauge potential may be divided into a geometric part $\vec{A}^{\rm G} (\vec{R})$
coming from the arbitrariness in the choice of the phases in the degenerate eigen-subspace
of the Hamiltonian and the dynamical part $\vec{A}^{\rm D} (\vec{R})$ such that
\begin{eqnarray}
\vec{A} (\vec{R}) = \vec{A}^{\rm G} (\vec{R})+ \vec{A}^{\rm D} (\vec{R}), \label{gaug dec}
\end{eqnarray}
where
\begin{eqnarray}
\vec{A}^{\rm D}_{j \alpha k \beta} (\vec{R}) = i \frac{\langle k \beta (\vec{R}) \vert \nabla_{\vec{R}} H(\vec{R}) \vert j \alpha (\vec{R}) \rangle}{e_j (\vec{R})-e_k (\vec{R})}. \label{gaug geom}
\end{eqnarray}
Here and hereafter, the superscript $D$ is used for the dynamical part in this section while
the subscript $D$ was and will be used for the diagonal matrix parts throughout the paper, which should
cause no confusion. The geometric part $\vec{A}^{\rm G} (\vec{R})$ cannot be determined from
the eigenvalue problem in Eq. (\ref{eigen st2}) and is the nonabelian effective gauge potential \cite{wilczek-zee}
also known as the Berry connection \cite{berry84}.

The transport of the frame of eigenstates along a curve $C(\vec{R})$ beginning at a base point $\vec{R}_0$
is given in the PI by
\begin{eqnarray}
\Phi_C (\vec{R}) = \prod_{C, \vec{R}_0}^{\vec{R}} \exp \bigl[- i \vec{A}_C (\vec{R}) \cdot d \vec{R} \bigr]
\Phi_C (\vec{R}_0). \label{transport}
\end{eqnarray}
The nonabelian gauge field tensor for a vector bundle of the gauge potential over the $\vec{R}$-space
follows from a commutator acting on the orthonormal frame (\ref{transport}) as
\begin{eqnarray}
\Bigl[ \frac{\partial}{\partial X_m}, \frac{\partial}{\partial X_n} \Bigr] \Phi_C (\vec{R})
= \Biggl(- i \Bigl( \frac{\partial A_{C, n} (\vec{R})}{\partial X_m} - \frac{\partial  A_{C, m} (\vec{R})}{\partial X_n} \Bigr)
- \bigl[A_{C, m} (\vec{R}), A_{C, n} (\vec{R}) \bigr] \Biggr) \Phi_C (\vec{R}) = F_{mn} (\vec{R}) \Phi_C (\vec{R}),
\label{field ten}
\end{eqnarray}
where the derivative $\partial/\partial X_m$ in the $\vec{R}$-space is actually a covariant derivative
in the horizontal subspace of the tangent space of the vector bundle \cite{EGH}.
Note that the nonabelian gauge field tensor $F_{mn} (\vec{R})$ for the induced
gauge potential in Eq. (\ref{gaug pot2}) satisfies the same field equation as a true gauge field; thus,
the field equation allows the interpretation of $\partial/\partial X_m$ as a covariant derivative.
For an abelian gauge potential, the commutator is the curl operator and can be used for the
(abelian) ordinary Stokes theorem. Equation (\ref{field ten}) relates the curl operator
to a gauge field tensor for a nonabelian gauge potential, too. The nonabelian Stokes
theorem was discussed in Ref. \cite{bralic}. However, in the history of the PI
(or matricant), the nonabelian Stokes theorem dates back to Schlesinger's formula \cite{schlesinger}
that relates the holonomy of the affine connection to the surface integral of the Riemannian-Christoffel
curvature tensor. The field tensor explains how the geometric phase for closed loops \cite{aharonov-anandan,page}
enters the evolution of the parameter-dependent quantum system.

The holonomy matrix in the PI, Eq. (\ref{transport}), may be evaluated by using the Magnus expansion \cite{magnus}:
\begin{eqnarray}
\prod_{C (\vec{R})} \exp \Bigl[- i \vec{A}_C (\vec{R}) \cdot d \vec{R} \Bigr] =
\exp \Bigl[ \int_{C(\vec{R})} M \bigl[ -i \vec{A}_C (\vec{R}) \bigr] \Bigr], \label{holon}
\end{eqnarray}
where the first term only with the geometric part is nothing but the Berry phase.
With the ordinary Stokes theorem applied, we see that
\begin{eqnarray}
\exp \Bigl[- i \int_{C} \vec{A}^{\rm G}_C (\vec{R}) \cdot d \vec{R} \Bigr] =
\exp \Bigl[- i \int_{S} \bigl(\nabla_{\vec{R}} \times \vec{A}^{\rm G}_C (\vec{R}) \bigr) \cdot d \vec{S} \Bigr]
= \exp \Bigl[ -i \int_{S} F^{\rm G}_C (\vec{R}) \cdot d \vec{S} \bigr] \Bigr],
\end{eqnarray}
where $d \vec{S}$ is an area vector for the surface bounded by $C(\vec{R})$ and $F^{\rm G}_C (\vec{R})$
is the Berry curvature in Eq. (\ref{field ten}) along that curve. Higher-order corrections to the Berry phase may be
calculated from Eq. (\ref{holon}) (see Appendix \ref{app-a} for higher-order terms for
the Magnus expansion.)

We now turn to Schlesinger's formula for a gauge field tensor. The holonomy matrices in Eq. (\ref{holon})
of the gauge potential for loops $C(\vec{R})$ at the base $\vec{R}_0$ form a holonomy group at $\vec{R}_0$.
Schlesinger's formula connects the holonomy of a gauge potential to a gauge field tensor.
After the affine connection and the Riemannian-Christoffel curvature tensor of differential geometry
are translated into the gauge potential and the gauge field tensor of gauge theory,
the holonomy in Eq. (\ref{holon}) is related to the gauge field tensor by a mixed product integral
\begin{eqnarray}
\prod_{C (\vec{R})} \exp \Bigl[- i \vec{A}_C (\vec{R}) \cdot d \vec{R} \Bigr] = \prod_{\phi^{-1} (S)}
\exp \Bigl[ \sum_{p < q} \int U_F (\xi, \eta) F_{qp} (\xi, \eta) U^{-1}_F (\xi, \eta)
J \Bigl(\frac{\phi_p, \phi_q}{\xi, \eta} \Bigr) d\xi d\eta \Bigr], \label{schles}
\end{eqnarray}
where $\phi_p$ is a surface map from a rectangle $\{ (x,y) \vert a \leq x \leq b, c \leq y \leq d \}$
in a Cartesian coordinate system to the loop $C(\vec{R})$ and its interior in the $\vec{R}$-space,
$J \bigl(\phi_p, \phi_q/ \xi, \eta \bigr)$ is the corresponding Jacobian, and
\begin{eqnarray}
U_F (\xi, \eta) = \prod_{b}^{\eta} \exp \Bigl[ - i A_m (\vec{R} (a, y)) \frac{\partial \phi_m (a,y)}{\partial y} dy \Bigr]
\prod_{a}^{\xi} \exp \Bigl[ - i A_n (\vec{R} (x, \eta)) \frac{\partial \phi_n (x,\eta)}{\partial x} dx \Bigr].
\end{eqnarray}
Schlesinger's formula in Eq. (\ref{schles}) is an exact one connecting the holonomy of
a gauge potential to a gauge field tensor. In fact, the Magnus expansion for the holonomy in Eq. (\ref{holon})
is one of the expansion methods for Schlesinger's formula.
Using some of the above results obtained in Secs. \ref{sec-II} and \ref{sec-III}, we shall develop
the superadiabatic expansion in the next section, and for that purpose a time-dependent Hamiltonian
with an adiabatic small parameter will be quite illustrative.

\section{Superadiabatic Expansion in the Nondegenerate Case}\label{sec-V}

In this and the next section, we shall reformulate the quantum evolution of a time-dependent
system in the PI. Then, the parameter is the explicit time, and the parameter space is a real axis or
a complex plane of time. The advantages of the PI formulation are the existence of useful properties and
the efficiency of the PI as a computational tool.

In order to develop the superadiabatic expansion, a generalization of the adiabatic expansion,
we shall insert explicitly an adiabatic parameter into the Schr\"{o}dinger equation for a time-dependent Hamiltonian:
\begin{eqnarray}
i \epsilon \frac{\partial}{\partial \tau} \vert \psi(\tau) \rangle = H(\tau) \vert \psi(\tau) \rangle, \label{sch eq}
\end{eqnarray}
where $\epsilon$ is an adiabatic small parameter depending on the Planck constant $\hbar$. The
quantum evolution operator for Eq. (\ref{sch eq}), which evolves a quantum state at a time
$\tau_0$ to a quantum state at an arbitrary time $\tau$, may be given in the PI by using
\begin{eqnarray}
\vert \psi(\tau) \rangle = \prod_{\tau_0}^{\tau} \exp \Bigl[- \frac{i}{\epsilon}
H(\tau) d\tau \Bigr] \vert \psi(\tau_0) \rangle.
\end{eqnarray}
In general, when expressed in the basis Eq. (\ref{eigen st2}) for $H(\tau)$, the PI becomes
\begin{eqnarray}
\prod_{\tau_0}^{\tau} \exp \Bigl[- \frac{i}{\epsilon}
H(\tau) d\tau \Bigr] = \Phi^T (\tau) \prod_{\tau_0}^{\tau} \exp \Bigl[-i \Bigl( \frac{1}{\epsilon}
H_D (\tau) - A^T (\tau) \Bigr) d\tau \Bigr] \Phi^*(\tau_0).
\end{eqnarray}
We can apply the general perturbation formula (sum rule) \cite{dollard-friedman} to factor out the dynamical phases:
\begin{eqnarray}
\prod_{\tau_0}^{\tau} \exp \Bigl[-i \Bigl( \frac{1}{\epsilon}
H_D (\tau) - A^T (\tau) \Bigr) d\tau \Bigr]= \exp \Bigl[-  \frac{i}{\epsilon} \int_{\tau_0}^{\tau}
H_{D} (\tau) d\tau \Bigr] \prod_{\tau_0}^{\tau} \exp \Bigl[i A_I (\tau) d\tau \Bigr],
\end{eqnarray}
where
\begin{eqnarray}
A_I (\tau) = \exp \Bigl[  \frac{i}{\epsilon} \int_{\tau_0}^{\tau} H_{D} (\tau) d\tau \Bigr]
A^{T} (\tau) \exp \Bigl[- \frac{i}{\epsilon} \int_{\tau_0}^{\tau} H_{D} (\tau) d\tau \Bigr].
\end{eqnarray}

Next, we shall use the formal equivalence between the PI and the time-ordered $T$-exponential \cite{kim91,kim92,magnus},
\begin{eqnarray}
 \prod_{\tau_0}^{\tau} \exp \Bigl[ i A_I (\tau) d\tau \Bigr] = I + \sum_{n=1}^{\infty} i^n
\int_{\tau_0}^{\tau} A_{I} (\tau') d \tau' \int_{\tau_0}^{\tau'} A_{I} (\tau^{''}) d \tau^{''}
\cdots \int_{\tau_0}^{\tau^{(n-1)}} A_{I} (\tau^{(n)}) d \tau^{(n)}, \label{T-exp}
\end{eqnarray}
and integrate by parts the rapidly oscillating factors in the adiabatic regimes to obtain
\begin{eqnarray}
 \prod_{\tau_0}^{\tau} \exp \Bigl[i A_I (\tau) d\tau \Bigr] = I + \sum_{n=1}^{\infty} \epsilon^n A_{I, n} (\tau, \tau_0).
\end{eqnarray}
The termination of the series in powers of the smallness parameter $\epsilon$ at the $N$th step gives another derivation
(not exactly the same) of Berry's superadiabatic basis:
 \begin{eqnarray}
\vert \psi_N (\tau) \rangle = \Phi^T (\tau) \exp \Bigl[-  \frac{i}{\epsilon} \int_{\tau_0}^{\tau} H_{D} (\tau) d\tau \Bigr]
\Bigl( I + \sum_{n=1}^{N} \epsilon^n A_{I, n} (\tau, \tau_0) \Bigr)  \Phi^*(\tau_0) \vert \psi (\tau_0) \rangle. \label{superad bas}
\end{eqnarray}
It is worth noting that the superadiabatic expansion is well-defined and converges sufficiently in the
adiabatic region in contrast to Berry's diverging basis for large $N$. Also, note that the superadiabatic
basis in Eq. (\ref{superad bas}) resulted from the $T$-exponential in Eq. (\ref{T-exp}), but this is not the only possible
form for the PI. For example, the formal solution in the Magnus expansion (Appendix \ref{app-a}) may be employed
to give the exponential correction
 \begin{eqnarray}
\vert \psi (\tau) \rangle = \Phi^T (\tau) \exp \Bigl[-  \frac{i}{\epsilon} \int_{\tau_0}^{\tau} H_{D} (\tau) d\tau \Bigr]
\exp \Bigl[ \sum_{n=1}^{\infty} \epsilon^n \tilde{A}_{I, n} (\tau, \tau_0) \Bigr]  \Phi^*(\tau_0) \vert \psi (\tau_0) \rangle \label{superad bas2}
\end{eqnarray}
and the $N$th truncation
 \begin{eqnarray}
\vert \tilde{\psi}_N (\tau) \rangle = \Phi^T (\tau) \exp \Bigl[-  \frac{i}{\epsilon} \int_{\tau_0}^{\tau} H_{D} (\tau) d\tau \Bigr]
\exp \Bigl[ \sum_{n=1}^{N} \epsilon^n \tilde{A}_{I, n} (\tau, \tau_0) \Bigr] \Phi^*(\tau_0) \vert \psi (\tau_0) \rangle.
\end{eqnarray}
Especially, when the group structure of the Hamiltonian operator is known, the Magnus expansion proves quite useful.
Now, we turn to the case when there are complex degenerate points coming from level crossings.

\section{Geometric Transition Amplitude}\label{sec-VI}

Let us suppose that the Hamiltonian $H(\tau)$ has an analytical continuation in a strip of the complex $\tau$-plane, denoted
by $z$, including the real $\tau$-axis, such that $H(\tau)$ has a basis of eigenstates with discrete eigenvalues except at
finite isolated complex points of the degeneracies. These conditions were used to derive the geometric transition
amplitude for a two-level system, and the applicability to general systems was suggested in Refs.
\cite{hwang-pechukas,MSW,berry90,JKP}. In this section, we shall reformulate the geometric transition amplitude
formula in a general form by using the analytic properties of the PI.

If there are no complex degenerate points for the extended Hamiltonian $H(z)$, then the path independence
of the quantum evolution operator is a consequence of the identity operator for the PI for a closed loop:
\begin{eqnarray}
\prod_{C_{I} (z), \tau_0} \exp \Bigl[-i \Bigl( \frac{1}{\epsilon}
H_D(z) - A^T (z) \Bigr) dz \Bigr] = \prod_{C_{II} (z), \tau_0} \exp \Bigl[-i \Bigl( \frac{1}{\epsilon}
H_D(z) - A^T (z) \Bigr) dz \Bigr]
\end{eqnarray}
for two different paths in the $z$-plane with the same initial and final points on the real axis. This
path independence makes the use of the complex $z$-plane legitimate. However, the situation is not
so simple when $H(z)$ has degeneracies at isolated complex points because $A^T (z)$ becomes
a meromorphic operator, whose regular singular behavior can be seen from Eqs. (\ref{gaug dec}) and
(\ref{gaug geom}) as the two or more energy eigenvalues approach each other. For the sake of simplicity,
we shall assume that a pair of complex conjugate points $z_0$ and $z_0^*$ are the only degenerate points.
Then, any curve connecting the initial and the final points is characterized by the homotopy class of winding
numbers around $z_0$ or $z_0^*$ or both $z_0$ and $z_0^*$. We shall restrict the study to the homotopy class
of the winding number around $z_0$, but the choice of the homotopy class should be determined from the preference
of exponentially small transition amplitudes for the superadiabatic expansion.

The quantum evolution operator for a curve $C(z)$ consisting of $C_0 (z)$ and $C_1^{(n)} (z)$ of winding number $n$,
($C(z) = C_0 (z)+C_1^{(n)} (z) $), is
\begin{eqnarray}
\prod_{C (z), \tau_0}^{\tau} \exp \Bigl[- \frac{i}{\epsilon}
\hat{H} (z) dz \Bigr] = \prod_{C_{0} (z), \tau_0}^{\tau} \exp \Bigl[- \frac{i}{\epsilon}
\hat{H} (z) dz \Bigr] \prod_{C_{1}^{(n)} (z_0)} \exp \Bigl[- \frac{i}{\epsilon}
\hat{H} (z) dz \Bigr],
\end{eqnarray}
where $\hat{H} (z) = H_{D} (z) - \epsilon A^T (z)$.
The PI's for two different closed loops with the same winding number are equivalent up to
a unitary transformation \cite{dollard-friedman}
\begin{eqnarray}
\prod_{C_{1}^{(n)} (z_0)} \exp \Bigl[- \frac{i}{\epsilon}
\hat{H} (z) dz \Bigr] = U \prod_{C_{1}^{'(n)} (z_0)} \exp \Bigl[- \frac{i}{\epsilon}
\hat{H} (z) dz \Bigr] U^{\dagger} \label{hom unit}
\end{eqnarray}
for some unitary matrix $U$. The unitary transformation equivalence of the ordered loop integrals
for the winding number one was also discussed in Ref. \cite{JKP}.

It is remarkable that the quantum evolution operator of the Hamiltonian with
complex degenerate points has an additional factor determined entirely by the PI
over the loops according to the homotopy class of degenerate points, which
is geometric in nature. We should note that our derivation of the
geometric factor was based only on the extension of the Hamiltonian to a strip
of the complex time plane including the real time axis and on the existence of isolated complex degenerate points.

The PI of meromorphic operators over loops beyond the simple poles is technically difficult
to perform, but a method may be put forward to replace the PI by the Magnus expansion (see Appendix \ref{app-a}).
Thus, the PI over the loops becomes
\begin{eqnarray}
\prod_{C_{1}^{(n)} (z_0)} \exp \Bigl[- \frac{i}{\epsilon}
\hat{H} (z) dz \Bigr] = \exp \Bigl[ \int_{{C_{1}^{(n)} (z_0)}} M \bigl[ - \frac{i}{\epsilon}
\hat{H}(z) \bigr] dz \Bigr], \label{holon mag}
\end{eqnarray}
where the Magnus global expansion up to the third order is
\begin{eqnarray}
M \bigl[ - \frac{i}{\epsilon} \hat{H}(z) \bigr] &=& - \frac{i}{\epsilon} \hat{H}(z) +
\frac{1}{2 \epsilon^2} \int_{z_0}^{z} dz' [\hat{H}(z'), \hat{H} (z)] +
\frac{i}{12 \epsilon^3} \int_{z_0}^{z} dz' \int_{z_0}^{z^{'}} dz^{''} [\hat{H}(z'), [\hat{H} (z^{''}), \hat{H} (z)]]
\nonumber\\ && + \frac{i}{4 \epsilon^3} \int_{z_0}^{z} dz' \int_{z_0}^{z^{'}} dz^{''} [[\hat{H} (z^{''}), \hat{H}(z')], \hat{H} (z)]]
+ \cdots.
\end{eqnarray}
The first term on the right-hand side yields
\begin{eqnarray}
\exp \Bigl[ - \frac{i}{\epsilon} \int_{{C_{1}^{(n)} (z_0)}}
\hat{H}(z) dz \Bigr] = \exp \Bigl[\frac{2 \pi n}{\epsilon} \hat{H} (z_0) \Bigr], \label{res geom}
\end{eqnarray}
where $\hat{H} (z_0)$ denotes the residue operator at $z_0$. The geometric factor in Eq. (\ref{res geom}) gives a generalization
of the Dykhne formula for the transition amplitudes for the homotopy class. Equation (\ref{holon mag}) includes higher-order corrections
for the geometric transition amplitudes. A complex curve starting on the real axis and ending on the real axis
can be homotopically deformed into a loop around one of the complex degenerate points and a segment of the real axis if the
curve goes initially outside them. The evolution operator is thereby composed of that of the real axis and a loop integral with
the winding number one around that complex degenerate point. The loop integrals with the winding number one are discussed
in the literature \cite{JKP}. The PI approach in this section is more systematic, and results thus obtained are more comprehensive
than others, leaving the investigation of all homotopy classes and the choice of paths ensuring the convergence of
the evolution operator for future research.

\section{Conclusion and Discussion}

The matrix effective Hamiltonian and the matrix effective Lagrangian were constructed in the product integral.
The matrix-operator approach in the product integral formulation provided a unified picture to quantum
evolution operators for parameter-dependent quantum systems. In the matrix effective Hamiltonian approach,
the Schr\"{o}dinger equation becomes a matrix equation, Eq. (\ref{MEH3}), under the unitary
transformation in Eq. (\ref{PI}). The matrix equation can be solved to find the energy eigenvalues
and the wave functions. As an application of the matrix effective Hamiltonian, a system of
a heavy particle coupled to a light particle of a harmonic oscillator, Eq. (\ref{mod ham}), was considered.
The two-component wave function in Eq. (\ref{har two-com}) and the superadiabatic expansion
in Eq. (\ref{har superad ex}) were obtained in the product integral. The superadiabatic expansion can give
asymptotic forms of the wave functions.

On the other hand, in the matrix effective Lagrangian approach, the full time-evolution operator kernel in Eq. (\ref{ker spec2})
and the matrix effective action in Eq. (\ref{kubo eff act}) were defined, and the Magnus expansion and the
Kubo cumulant expansion gave systematically the higher-order corrections to the adiabatic kernel and
the effective action. As a by-product, we found that up to the lowest Magnus expansion and the
Kubo cumulant expansion, the effective action in Eq. (\ref{har eff act2}) for the system in Eq. (\ref{har lag})
of a heavy particle coupled to a light particle of a harmonic oscillator involves the new (mode-dependent)
gauge potential in Eq. (\ref{har eff gaug}) that has been known not to exist for the nondegenerate case (real eigenfunctions).
The real eigenfunctions, Eq. (\ref{har wav}), were chosen, and the induced gauge potential, Eq. (\ref{har eff gaug}),
has its origin in a sequence of transitions accompanying the creation of two quanta, followed by the annihilation
of two quanta or vice versa, in the number-state representation.

The holonomy, Eq. (\ref{holon}), of the gauge potential (Berry connection) could be evaluated using the Magnus expansion,
and the holonomy was observed to be related to Schlesinger's formula for the gauge field tensor in Eq.
(\ref{schles}). Schlesinger's formula gave the exact geometric phase. The geometric transition
amplitudes for the quantum evolution operators were also formulated in the product integral and found
to depend only on the contour integrals of a homotopy class around complex degenerate points.

The matrix-operator approach to quantum systems, which seemed especially applicable to
parameter-dependent quantum systems, may provide an alternative method to address
further both the canonical quantum field theory and the path integral formalism.
Moreover, the matrix approach is nonperturbative as far as the spectra resolution
can be completely employed and the product integral  of matrix operator can be
found in a closed form. However, the Magnus formal solution (expansion)
for the product integral provides at least a perturbation method,
which appears quite different from the conventional ones
in that the first term in the Magnus expansion is the leading term as in Sec. \ref{sec-III}.
The matrix effective Hamiltonian approach might be applied to canonical
quantum gravity, whose wave function is a generalization of the adiabatic wave function
to a time-dependent Klein-Gordon-like equation. It
may be used to find the complete set of solutions for such systems and to provide
the Hilbert space structure to the set of solutions \cite{kim92}. The meaning
of the new (mode-dependent) gauge potential for quantum gravity is left for
a future investigation.

The matrix effective Lagrangian approach might be applicable
to finding systematically the back-reaction of subsystems. The higher-order back-reactions
are expected to contribute small amounts to the lower modes in the adiabatic region,
in which the matrix effective Lagrangian method is efficient, but may contribute
significant amounts to the higher modes. In other cases such as level crossings where
the adiabatic method fails to work, the matrix-valued gauge potential
contains a regular singular part, and the matrix effective Lagrangian method in
Sec. \ref{sec-III} might not work. One needs to treat the singular part carefully and
may expect geometric factors to enter in the path integral, depending on the contours
of the homotopy class of the simple poles of the matrix-valued gauge potential as in Sec. \ref{sec-VI}.
The singular gauge potential introduces a geometric factor for the quantum evolution operator
in Sec. \ref{sec-VI}. The quantum theory of vortices in nonabelian gauge theory will provide
such a challenging problem.

\medskip

\noindent{\bf Note added}: This paper was an unpublished report (SNUTP-92-10) to the Center for Theoretical Physics, Seoul National University, in 1992.
The only revision in this paper is the reinterpretation of Eq. (\ref{har eff act2}).
After completion of this paper, the extension of the Born-Oppenheimer equation was elaborated
in Refs. \cite{mead,baer1,baer2,sardanashvily,viennot}. The adiabatic theorem has also been studied in Refs.
\cite{marzlin-sanders,TSKO,wu-yang,MMPP,zhao,fujikawa08,yukalov,ROP,comparat,amin,boixo-somma,tong}.
Another matrix-operator method for the effective action
has been developed for the minimal scalar field WDW equation, in which the
Born-Oppenheimer method separates the gravitational part from the scalar field and the pilot-wave theory along the peaked wave
packets leads to the semiclassical gravity and a unitary quantum field theory \cite{kim95}.
The Magnus expansion has been elaborated in Ref. \cite{BCOR}.

\acknowledgments
SPK would like to thank Misao Sasaki and Takahiro Tanaka for the warm hospitality at
the Yukawa Institute for Theoretical Physics at Kyoto University,
where this work was revised during the Long-term Workshop YITP-T-12-03 on ``Gravity and Cosmology 2012,''
and Pauchy W-Y. Hwang for the warm hospitality at National Taiwan University, where the revision was completed.
The revision of this paper was supported in part by the Basic Science Research Program through
the National Research Foundation of Korea (NRF) funded by the Ministry of Education, Science and
Technology (NRF-2012R1A1B3002852).

\appendix

\section{Magnus Formal Solution (Expansion)} \label{app-a}

We briefly summarize the Magnus formal solution (expansion) to the PI whose detailed description may be found in the Appendix of Refs.
\cite{kim92,BCOR}. An operator-valued first-order linear differential (evolution) equation
\begin{eqnarray}
\frac{dU(t)}{dt} = O(t) U(t) \label{a-1}
\end{eqnarray}
has a formal solution in the PI \cite{dollard-friedman}:
\begin{eqnarray}
U(t) \equiv \prod_{t_0}^t \exp \bigl[ O(t') dt' \bigr] U(t_0). \label{a-2}
\end{eqnarray}
Thus, the PI is equal to the time-ordered integral
\begin{eqnarray}
\prod_{t_0}^t \exp \bigl[ O(t') dt' \bigr] = {\cal T} \exp \Bigl[ \int_{t_0}^t O(t') dt' \Bigr]. \label{pi-ti eq}
\end{eqnarray}
The PI may be found in a closed form in a certain case; otherwise, we should rely on the formal (perturbative)
solution, known as the Magnus expansion \cite{magnus}. The PI can be written in the Magnus expansion
\begin{eqnarray}
\prod_{t_0}^t \exp \bigl[ O(t') dt' \bigr] = \exp \bigl[ \int_{t_0}^t M[O(t')] dt' \bigr]. \label{a-3}
\end{eqnarray}
The commutator approach to the Magnus expansion follows from equating the derivatives of both sides of Eq. (\ref{a-3}):
\begin{eqnarray}
M = \sum_{n=0}^{\infty} \frac{B_n}{n!} \Bigl({\rm ad} \int_{t_0}^t M[O(t')] dt' \Bigr)^n O(t), \label{a-4}
\end{eqnarray}
where $B_n$ are the Bernoulli numbers. Then, the Magnus expansion can be found as an infinite sum \cite{kim92}:
\begin{eqnarray}
M = \sum_{n_1=0}^{\infty} S^{*}_{n_1} \Biggl({\rm ad} \int_{t_0}^t  \sum_{n_2=0}^{\infty} S^{*}_{n_2} \Biggl({\rm ad}
 \int_{t_0}^{t_1} \sum_{n_3=0}^{\infty} S^{*}_{n_3} \Biggl({\rm ad} \int_{t_0}^{t_2} ( \cdots) \Biggr)^{n_4} O(t_4) \Biggr)^{n_3} O(t_3)\Biggr)^{n_2} O(t_2) \Biggr)^{n_1} O(t_1), \label{a-5}
\end{eqnarray}
where $S^{*}_n = B_n /n!$ and
\begin{eqnarray}
\bigl({\rm ad} A \bigr)^n B = [A, \bigl({\rm ad} A \bigr)^{n-1} B], \quad \bigl({\rm ad} A \bigr)^0 B = B. \label{a-6}
\end{eqnarray}

Here, the integrations are ordered from the inside to the outside. One may insert an expansion
parameter such as $1/h$ to the operator $O$ and perform the multi-power expansions and the multi-partitions
of integers into nonnegative integers. The first few terms of the Magnus expansion are
\begin{eqnarray}
M_{(0)} &=& O, \nonumber\\
M_{(1)} &=& - \frac{1}{2} \Bigl( {\rm ad} \int O \Bigr) O, \nonumber\\
M_{(2)} &=& \Bigl(- \frac{1}{2} \Bigr)^2 \Bigl( {\rm ad} \int \Bigl( {\rm ad} \int O \Bigr) O \Bigr) O
+ \frac{1}{12} \Bigl( {\rm ad} \int O \Bigr)^2 O, \nonumber\\
M_{(3)} &=& \Bigl(- \frac{1}{2} \Bigr)^3 \Bigl( {\rm ad} \int \Bigl( {\rm ad} \int \Bigl( {\rm ad} \int O \Bigr) O \Bigr) O \Bigr) O
+ \Bigl( - \frac{1}{2} \Bigr) \frac{1}{12} \Bigl( {\rm ad} \int \Bigl( {\rm ad} \int O \Bigr)^2 O \Bigr) O. \label{a-7}
\end{eqnarray}
The Magnus formal solution (expansion) Eq. (\ref{a-5}) can be used for the MEL in Sec. {\ref{sec-III} and
the exponential form of the superadiabatic basis in Sec. \ref{sec-IV} because the commutators may get rid of
large terms.

Other methods, such as Burum's method \cite{burum} and Salzman's method \cite{salzman}, follow
from the formal equivalence Eq. (\ref{pi-ti eq}) between the PI and the time-ordered integral. The Magnus expansion in these methods
is an exponential operator in a series of combinations of time-ordered integrals. These methods might be used for
calculating the higher-order geometric phase Eq. (\ref{holon}) in Sec. \ref{sec-IV}.

We now list some useful properties used in this paper. The general perturbation formula (sum rule) is
\begin{eqnarray}
\prod_{t_0}^t \exp \bigl[ (O_I (t')+O_{II} (t')) dt' \bigr] = P(t) \prod_{t_0}^t \exp \bigl[ P^{-1} (t') O_{II} (t') P(t') dt' \bigr], \label{a-8}
\end{eqnarray}
where
\begin{eqnarray}
P(t) = \prod_{t_0}^t \exp \bigl[ O_I (t') dt' \bigr]. \label{a-9}
\end{eqnarray}
The similarity rule for an invertible operator $M (t)$ is
\begin{eqnarray}
\prod_{t_0}^t \exp \bigl[ O (t')  dt' \bigr] = M(t) \prod_{t_0}^t \exp \Bigl[\Bigl( M^{-1} (t') O (t') M(t') - M^{-1} (t')
\frac{dM(t')}{dt'} \ \Bigr)dt' \Bigr] M^{-1} (t_0). \label{a-10}
\end{eqnarray}
The similarity rule resembles the diagonalization of the matrix (similarity transformation for a symmetric matrix) in a certain sense.

\section{Kubo Cumulant Expansion} \label{app-c}

We summarize the first few terms of the cumulant expansion \cite{kubo}:
\begin{eqnarray}
\langle O \rangle_c &=& \langle O \rangle, \nonumber\\
\langle O^2 \rangle_c &=& \langle O^2 \rangle - \langle O \rangle^2, \nonumber\\
\langle O^3 \rangle_c &=& \langle O^3 \rangle - 3 \langle O \rangle \langle O^2 \rangle + 2 \langle O \rangle^3, \nonumber\\
\langle O^4 \rangle_c &=& \langle O^4 \rangle - 4 \langle O \rangle \langle O^3 \rangle - 3
\langle O \rangle^2 \langle O^2 \rangle + 12 \langle O \rangle^3 \langle O \rangle - 6 \langle O \rangle^4. \label{c-1}
\end{eqnarray}

\end{document}